\documentclass[showpacs,preprintnumbers,amssymb]{revtex4-2}

\AtBeginDocument{
  \setlength\abovedisplayskip{0pt}
  \setlength\belowdisplayskip{0pt}}

\usepackage{graphicx}
\usepackage{dcolumn}
\usepackage[dvipsnames]{xcolor}
\usepackage{bm}
\usepackage{amsmath}
\usepackage{amssymb}
\usepackage{epsfig}
\usepackage{amsfonts}
\usepackage{lineno,hyperref}
\usepackage{array}
\usepackage{float}
\usepackage{microtype}
\usepackage{multirow}
\usepackage{adjustbox}
\usepackage[english]{babel}
\usepackage{epstopdf}
\usepackage{blindtext}
\usepackage{subcaption}

\def \a{\alpha}
\def \b{\beta}
\def \l{\lambda}

\def \k{\kappa}

\def \p{\partial}

\def \f{\phi}

\def \be{\begin{equation}}
\def \ee{\end{equation}}
\def \ben{\begin{eqnarray}}
\def \een{\end{eqnarray}}

\def \G{\bar{G}}
\def \r{\tilde{r}}
\def \rh{\bar{\rho}}
\def \K{\dot{\phi}^2}
\def \R{\bar{R}}

\def \half{\frac{1}{2}}

\def \T{\bar{T}}
\def \Th{\bar{\Theta}}
\def \La{\mathcal{L}}
\def \F{\mathcal{F}}

\DeclareUnicodeCharacter{2212}{-}
\begin{document}

\title{Thermodynamics of a Non-canonical $f(\bar{R},\bar{T})$ gravity}   

\author{Arijit Panda}
\email{arijitpanda260195@gmail.com}
\affiliation{Department of Physics, Raiganj University, Raiganj, Uttar Dinajpur, West Bengal, India, 733 134. $\&$\\
Department of Physics, Prabhat Kumar College, Contai, Purba Medinipur, India, 721 404.}
\author{Goutam Manna$^a$}
\email{goutammanna.pkc@gmail.com}
\altaffiliation{$^a$Corresponding author}
\affiliation{Department of Physics, Prabhat Kumar College, Contai, Purba Medinipur 721404, India $\&$\\ Institute of Astronomy Space and Earth Science, Kolkata 700054, India}
\author{Saibal Ray}
\email{saibal.ray@gla.ac.in}
\affiliation{{Centre for Cosmology, Astrophysics and Space Science (CCASS), GLA University, Mathura 281406, Uttar Pradesh, India}}

\author{Maxim Khlopov}
\email{khlopov@apc.in2p3.fr}
\affiliation{Institute of Physics, Southern Federal University, 194 Stachki, Rostov-on-Don 344090, Russian Federation $\&$\\
National Research Nuclear University, MEPHI, Moscow, Russian Federation $\&$\\ Virtual Institute of Astroparticle Physics 10, rue Garreau, 75018 Paris, France}
\author{Praveen Kumar Dhankar}
\email{pkumar6743@gmail.com}
\affiliation{Symbiosis Institute of Technology, Nagpur Campus, Symbiosis International (Deemed University), Pune-440008, India}

\begin{abstract}
This work comprises a study of the thermodynamic behavior of modified $f(\bar{R},\bar{T})$ gravity, which had been developed based on a non-canonical theory known as K-essence theory. In this development, we use the Dirac-Born-Infeld (DBI) type of non-standard Lagrangian. We develop a modified first law and generalized second law of thermodynamics (GSLT) within the modified $f(\bar{R},\bar{T})$ gravity, where we consider the background metric to be the usual Friedmann-Lema$\hat{\text{i}}$tre-Robertson-Walker (FLRW) type.  A graphical analysis of surface gravity has been performed for the modified FLRW metric via the $f(\bar{R},\bar{T})$ theory, which is different from the usual FLRW gravity through the usual $f(R,T)$ gravity. Exponential and power law scale factors are used to analyze cosmic surface gravity. Through the investigation of the modified GSLT, using the relation of scale factor with the scalar field, we have seen that during the initial phase of the universe, the entropy's rate of change may be either negative or positive, contingent upon the value of the curvature constant. The negativity of the entropy change indicates that the modified GSLT is not feasible in that particular area for a particular curvature constant. These traits suggest that during the inflationary period, entropy might have been either negative or positive. It has also been seen that entropy saturates every curvature value at different time ranges, which indicates the heat death of the universe.
\end{abstract}

\keywords
{Modified theories of gravity, Thermodynamics, Dark components, Entropy, K-essence
} 

\pacs{04.20.-q, 04.20.Cv, 04.50.Kd, 98.80.-k}

\maketitle



\section{Introduction}
Some cosmological queries, such as the horizon, the monopole, and the flatness problem, remained unanswered by the standard Einstein's gravitational theory. Furthermore, the fine-tuning problem ~\cite{Yoo} requires scientists to modify the theory due to the significant disparity in energy density, which is of the order of $10^{120}$. In addition, the `Cosmic Coincidence Problem' ~\cite{Velten,Yoo} has nevertheless to be resolved within the framework of the standard $\Lambda$-Cold Dark Matter (henceforth $\Lambda$CDM) theory. The recent observational data ~\cite{Perlmutter,Hawkins,SDSS} implies the universe is driven by an exotic component called dark energy by almost $70\%$. In order to address the foregoing issues, a new modified theory was proposed under the name of $f(R)$ theory~\cite{Nojiri_2006ri,Nojiri_2017ncd,Nojiri_2010wj,Sotiriou}. The Einstein-Hilbert (EH) action was considered to be contingent on the function of the Ricci scalar ($f(R)$) rather than merely $R$. Harko et al. \cite{Harko} expanded upon this theory by integrating the trace of the energy-momentum tensor $T$ into the function, resulting in the $f(R,T)$ theory. Later on, the authors of \cite{Barrientos} corrected the conservation equation of \cite{Harko}. Over the period of the last several years, a number of fascinating works have been conducted in relation to this theory ~\cite{Barrientos2,Carvalho1,Carvalho2,Ordines,Panda1,Panda2,Bouali,Tiago2,Panda4}.

Meanwhile, in order to address the cosmological unsolved problems mentioned earlier, scientists have put forth various models, including non-canonical ones like K-essence. The K-essence theory, which is shown in~\cite{Picon3,Picon1,Picon2,Scherrer,Chimento2,Visser,Babichev2,Vikman,Mukohyama,gm1,gm2,gm3}, is not the same as the usual relativistic scalar field theory. As demonstrated in \cite{Goldstein, Rana, Raychaudhuri} the non-canonical Lagrangian is the general one that can produce canonical Lagrangian under specific circumstances.  The authors of~\cite{gm1,gm2,gm3} used the Dirac-Born-Infeld (DBI) type~\cite{Born,Born2,Dirac} non-standard Lagrangian to come up with the simplest form of the K-essence emergent metric $\bar{G}_{\mu\nu}$, which is not conformally equivalent to the usual gravitational metric $g_{\mu\nu}$. The non-canonical Lagrangian can have the form of $\La(X,\phi)=-V(\phi)F(X)$~\cite{Vikman,Picon1,Picon2} or $\La (X,\phi)=F(X)-V(\phi)$ ~\cite{Dutta,Santiago} or $\La (X,\phi)\equiv \La (X)= F(X)$ \cite{Scherrer,Mukohyama} where $F(X)\equiv \La(X) (\neq X)$ is the non-canonical kinetic part with $X=\frac{1}{2} g_{\mu\nu} \nabla^{\mu}\phi \nabla^{\nu} \phi$, $V(\phi)$ is the canonical potential part. The underlying motivation of using non-canonical theory has been well-described in~\cite{Panda3}.  We discuss two main advantages of K-essence theory here~\cite{Panda3}. Firstly, the radiation background is the only factor influencing the K-essence field's behavior in the K-essence model until the dark energy-dominated era started. It eliminates the requirement for fine-tuning that the canonical model had. Secondly, it has the potential to produce a dark energy component in which the speed of sound ($c_S$) is always slower than the speed of light. Large angular scale cosmic microwave background (CMB) disturbances may be reduced by this characteristic. It has also been said that the K-essence theory can be used in a model of dark energy \cite{gm1,gm2,gm3,gm6} as well as from a gravitational point of view \cite{gm4,gm7,gm8,gm9} without taking into account the dark parts of the universe. Moreover, K-essence theory can be utilised to study unified dark energy and dust dark matter \cite{Guendelman_2016}, in the field of inflation and dark energy \cite{Guendelman_2015}. Panda et al.~\cite{Panda3} investigated the cosmic behaviors of the cosmos using a non-canonical Lagrangian within the framework of emergent $f(\R,\T)$ gravity. The researchers used the Lagrangian of the DBI type, denoted as $\La(X)=1-\sqrt{1-2X}$, to derive the Friedmann equations. They then computed the effective energy density ($\rh$), pressure ($\bar{p}$) and the Equation of State (EoS) parameter ($\bar{\omega}$). In this regard, we would like to mention that unification of inflation and dark sector of the universe has also been done in other theories \cite{Guendelman_2022,Guendelman_2023}. The authors introduce a two-scalar field model using two-measure theory (TMT) to unify the early and present universe. In the Einstein frame, this model generates K-essence, leading to a non-singular emergent universe followed by inflation in the early universe, dark epochs with dark energy (DE), dark matter (DM) and stiff matter in the present universe. 

On the other hand, black hole thermodynamics reveals a captivating link between our empirical observations of black holes and the fundamental principles of thermodynamics. Black holes exhibit similarities to regular thermodynamic systems, despite being composed of empty spacetime. These properties involve entropy and temperature. The entropy of a black hole is tightly linked to the area of its event horizon, while the temperature is intimately connected to the intriguing phenomenon known as Hawking radiation. In simple terms, based on the principles of quantum physics and general relativity, a black hole can be thought of as a black body that gives off thermal radiation. The temperature of this radiation is directly linked to the surface gravity at the black hole's horizon, while the entropy is tied to the area of the horizon. The Hawking temperature and horizon entropy, along with the black hole mass, follow the principles of the first law of thermodynamics ~\cite{Bardeen,Hawking,Bekenstein}. The entropy and temperature formulas for black holes show a certain degree of universality, which means that Einstein's equations alone determine the size of the black hole's horizon and the gravity of its surface. The finding of Bekenstein on black hole entropy provided evidence for a significant physical relationship between thermodynamics and gravity \cite{Bekenstein}. Based on Einstein gravity, the gravitational entropy $S$ relates to the black hole horizon area $A$ as $S = A/(4G)$, where $G$ is the gravitational constant and Hawking temperature $(T)$ is related to the surface gravity ($\k_{sg}$) as $T=\k_{sg}/2\pi$, which corresponds to the first law of thermodynamics \cite{Bekenstein, Bardeen, Hawking}. The Clausius relation $T dS = dQ$ on all local acceleration horizons in the Rindler space-time combined with the relation $S\propto A$ allows for the derivation of Einstein equations, as demonstrated by Jacobson ~\cite{Jacobson}. Here, $dQ$ and $T$ stand for the energy flux across the horizon and the Unruh temperature~\cite{Unruh} perceived by an accelerating observer just inside the horizon, respectively. This approach was used in several cosmological scenarios, such as the quasi-de Sitter inflationary universe~\cite{Frolov,Danielsson} and the dark energy-dominated realm~\cite{Bousso}. It is to note that Cai and Kim \cite{Cai} used the first law of thermodynamics to find the entropy of the apparent horizon in the Friedmann-Lema{\i'}tre-Robertson-Walker (FLRW) universe, taking into account any spatial curvature. On the other hand, later on Akbar and Cai~\cite{Akbar} showed that the Friedmann equations in GR may be stated as follows at the apparent horizon: $dE=TdS+WdV$, where $W=\half(\rho-p)$, $V$, $\rho$ and $p$ are respectively the work density, volume, energy density and pressure while $E = \rho V$ is the total energy within the apparent horizon.

Several modified theories of gravity have shown that thermodynamics and gravity are connected. These include scalar-tensor gravity~\cite{Cai,Bamba}, Lovelock gravity~\cite{Cai,Cai3}, Braneworld gravity~\cite{Sheykhi}, nonlinear gravity~\cite{Bamba1}, and Gauss-Bonnet gravity \cite{Akbar2} and in many more articles \cite{Bamba_2009gq,Odintsov_2024hzu,Nojiri_2023wzz,Nojiri_2022nmu,Volovik}. To find equilibrium thermodynamics of the apparent horizon in the expanding cosmological background, the authors of~\cite{Bamba} used a large group of modified gravity theories with the Lagrangian density $f(R,\phi,X)$. In these theories, $R$ is the Ricci scalar and $X$ is the kinetic part of a scalar field $\phi$. Note that non-equilibrium representation of thermodynamics is necessary in $f(R)$ gravity and scalar-tensor theory in order to modify the Clausius relation to $T dS = \delta Q + d\bar{S}$. The additional entropy production term in this case is $d\bar{S}$~\cite{Akbar,Bamba1,Sharif3}. In the study~\cite{Sharif3}, the authors establish the principles of thermodynamics in $f(R,T)$ gravity and demonstrate that the existence of a connection between matter and geometry, represented by the expression $f_{T}=\partial f(R,T)/\partial T$, prevents the attainment of an equilibrium state. In $f(R, T)$ gravity, the presence of the total energy exchange term ($q_{tot}$) indicates that there is an ongoing energy exchange that occurs with the horizon.  There is always a non-zero entropy-producing term in this theory. According to the authors, the non-equilibrium description may be recognized as a consequence of energy flow happening both inside and beyond the apparent horizon. The absence of an equilibrium state is caused by the non-zero value of the $f_{T}$ term. If the Lagrangian is only dependent on the geometry, then the term $f_{T}$ becomes zero, resulting in the attainment of an equilibrium state. This phenomenon occurs in the context of $f(R)$ gravity. Additionally, they demonstrated that both the phantom and non-phantom phases of $f(R, T)$ gravity adhere to the Generalised Second Law of Thermodynamics (GSLT). This result is consistent with the hypothesis put out by Nojiri et al. ~\cite{Nojiri} that entropy might have a positive value even in the phantom period.

The objective of this research is to analyze the thermodynamic properties of the universe within the framework of K-essence $f(\R,\T)$ gravity. The underlying metric used for our investigation is the homogeneous and isotropic FLRW spacetime in the context of K-essence. As noted in \cite{Panda4}, the question about the selection of the additive form of the $f(R,T)$ function (as discussed in \cite{Fisher,Harko2,Fisher2}), can be resolved in our scenario due to the non-canonical Lagrangian ($\La(X)$) possessing an indirect dependence on the metric. The benefits and applicability of K-essence theory and its relationship with $f(R,T)$ theory have been extensively discussed in \cite{Panda2,Panda4,Panda3}. The various cosmological scenarios have been successfully studied earlier \cite{Panda2,Panda4,Panda3}, which triggered us to use this theory in thermodynamics to check its consistency. We aim to find how the extra interactive terms in this non-canonical theory affect the laws of thermodynamics over GR or other modified theories. The second law of thermodynamics, which states that entropy always increases, plays a key role in the evolution of the universe. In the context of this modified gravity, we would like to study how the entropy of the universe evolves differently compared to GR. This will be definitely offering insights into the large-scale structure, cosmic acceleration and the thermodynamic arrow of time. Also, as we are working with a theory that fundamentally modifies the metric of the space-time, therefore the thermodynamical quantities may vary and show some different nature in the context of this theory.

This work involves six sections. The second section has been dedicated to a review of the non-standard theory of K-essence. Section III comprises a review of $f(\R,\T)$ gravity from the point of view of K-essence. In Section IV, we have studied the thermodynamics of $f(\R,\T)$ gravity. In particular, subsection IV-A is the formulation of the first law of thermodynamics and a comparative study of surface gravity between the usual FRW and modified FRW cases. Subsection IV-B derives the generalized second law of thermodynamics of the $f(\R,\T)$ gravity. In Section V, we focus on the study of first-order and second-order variations of entropy with time. Section VI is the conclusion.

\section{The K-essence}

This section provides a concise overview of K-essence theory. The K-essence geometry involves the interaction of a scalar field with gravity, which may be described by the following action when the scalar field is minimally coupled to gravity~\cite{Babichev2,Vikman,Visser,gm1,gm2,gm3}:
\ben
S_{k}[\phi,g_{\mu\nu}]= \int d^{4}x {\sqrt -g} \La(X,\phi),
\label{1}
\een
where $X=\frac{1}{2}g^{\mu\nu}\nabla_{\mu}\phi\nabla_{\nu}\phi$ is the canonical kinetic term and $\La(X,\phi)$ is the non-canonical Lagrangian. From the definition of the energy-momentum tensor:
\ben
T_{\mu\nu}\equiv \frac{-2}{\sqrt {-g}}\frac{\delta S_{k}}{\delta g^{\mu\nu}}=-2\frac{\partial \La}{\partial g^{\mu\nu}}+g_{\mu\nu}\La=-\La_{X}\nabla_{\mu}\phi\nabla_{\nu}\phi
+g_{\mu\nu}\La, 
\label{2}
\een
where $\La_{\mathrm X}= \frac{d\La}{dX},~\La_{\mathrm XX}= \frac{d^{2}\La}{dX^{2}}, ~\La_{\mathrm\phi}=\frac{d\La}{d\phi}$ and $\nabla_{\mu}$ is the covariant derivative defined with respect to the gravitational metric $g_{\mu\nu}$. 

The corresponding scalar field equation of motion (EOM) is
\ben
-\frac{1}{\sqrt {-g}}\frac{\delta S_{k}}{\delta \phi}= \tilde{G}^{\mu\nu}\nabla_{\mu}\nabla_{\nu}\phi +2X\La_{X\phi}-\La_{\phi}=0,
\label{3}
\een
where the effective metric is:
\ben
\tilde{G}^{\mu\nu}\equiv \frac{c_{s}}{\La_{X}^{2}}[\La_{X} g^{\mu\nu} + \La_{XX} \nabla ^{\mu}\phi\nabla^{\nu}\phi]
\label{4}
\een
with $1+ \frac{2X\La_{XX}}{\La_{X}} > 0$ and $c_s^{2}(X,\phi)\equiv{(1+2X\frac{\La_{XX}} {\La_{X}})^{-1}}$.

After a conformal transformation \cite{gm1,gm2,gm3} $\bar G_{\mu\nu}\equiv \frac{c_{s}}{\La_{X}}G_{\mu\nu}$, we can write the inverse of the effective metric (\ref{4}) as
\ben
\bar{G}_{\mu\nu}=g_{\mu\nu}-\frac{\La_{XX}}{\La_{X}+2X\La_{XX}}\nabla_{\mu}\phi\nabla_{\nu}\phi.
\label{5}
\een

Equations (\ref{3})--(\ref{5}) are physically relevant if $\La_{X}\neq 0$ is positive definite. In K-essence, Eq. (\ref{5}) states that our emergent metric, $\G_{\mu\nu}$, is conformally different from $g_{\mu\nu}$ (disformally related) for non-trivial spacetime configurations of $\f$. Unlike canonical scalar fields, $\f$ has different local causal structural features. They also vary from those specified with $g_{\mu\nu}$. The disformal transformation was thoroughly examined and analyzed by Bekenstein \cite{Bekenstein_1993}.

By addressing the Lagrangian's implicit dependency on $\f$, the equation of motion (EOM) expressed in Eq. (\ref{3}) becomes,
\ben
\frac{1}{\sqrt{-g}}\frac{\delta S_{k}}{\delta \phi}= \bar G^{\mu\nu}\nabla_{\mu}\nabla_{\nu}\phi=0.
\label{6}
\een

Taking into account the non-canonical Lagrangian of the DBI type $\La(X,\phi)\equiv \La(X)$~\cite{Mukohyama,Panda2,Panda3,Born,Born2,Dirac}:
\ben
\La(X)= 1-\sqrt{1-2X},
\label{7}
\een
we have the effective emergent metric (\ref{5}) transformed as
\ben
\bar G_{\mu\nu}= g_{\mu\nu} - \nabla_{\mu}\phi\nabla_{\nu}\phi= g_{\mu\nu} - \partial_{\mu}\phi\partial_{\nu}\phi,
\label{8}
\een
since $\phi$ is a scalar. In equation (\ref{7}), the potential part is eliminated due to the dominance of kinetic energy over potential energy in the K-essence geometry, as indicated by Mukohyama ~\cite{Mukohyama,Panda2, Panda3}. The value of $c_{s}^{2}$ is $(1-2X)$. 

Following~\cite{gm1,gm2}, the Christoffel symbol associated with the emergent gravity metric Eq. (\ref{9}) is: 
\ben
\bar\Gamma ^{\alpha}_{\mu\nu} 
&&=\Gamma ^{\alpha}_{\mu\nu} -\frac {1}{2(1-2X)}\Big[\delta^{\alpha}_{\mu}\partial_{\nu}
+ \delta^{\alpha}_{\nu}\partial_{\mu}\Big]X,~~~~~~~~~~~
\label{9}
\een
where $\Gamma ^{\alpha}_{\mu\nu}$ is the usual Christoffel symbol associated with the gravitational metric $g_{\mu\nu}$ and the corresponding geodesic equation for the K-essence geometry becomes:
\ben
\frac {d^{2}x^{\alpha}}{d\l^{2}} +  \bar\Gamma ^{\alpha}_{\mu\nu}\frac {dx^{\mu}}{d\l}\frac {dx^{\nu}}{d\l}=0, \label{10}
\een
where $\l$ is an affine parameter.

The literature~\cite{Babichev2,Panda3} describes the covariant derivative $D_{\mu}$, which is linked to the emergent metric $\bar{G}_ {\mu\nu}$ and meets the condition $D_{\a}\bar{G}^{\a\b}=0$. It can be expressed as
\ben
D_{\mu}A_{\nu}=\partial_{\mu} A_{\nu}-\bar \Gamma^{\l}_{\mu\nu}A_{\l}, \label{11}
\een
and the inverse emergent metric is $\bar G^{\mu\nu}$ such that $\bar G_{\mu\l}\bar G^{\l\nu}=\delta^{\nu}_{\mu}$.

It can be observed that the "Emergent Einstein's Field Equation (EEFE)" can be recast by considering the complete action that describes the dynamics of K-essence and general relativity \cite{Vikman, Panda3, Panda2} as
\ben
\bar{\mathcal{G}}_{\mu\nu}=\R_{\mu\nu}-\frac{1}{2}\bar{G}_{\mu\nu}\R=\k \T_{\mu\nu}, 
\label{12}
\een
where $\k=8\pi G$ is constant, $\R_{\mu\nu}$ is the Ricci tensor, $\R~ (=\R_{\mu\nu}\bar{G}^{\mu\nu})$ is the Ricci scalar and $\T_{\mu\nu}$ is the energy-momentum tensor of the K-essence geometry. Understanding the relationship between the EEFE and the K-essence geometry framework is essential. The equation of this geometry becomes exactly the same as the standard Einstein field equation when we remove the K-essence scalar field ($\f$). Given the current circumstances, it is plausible to argue that the Einstein field equation and our understanding of geometry may undergo modifications.

\section{$f(\R,\T)$ gravity via K-essence}

Under K-essence geometry, Panda et. al.~\cite{Panda3, Panda2} established the modified $f(\R, \T)$ gravity. The action of the $f(\R,\T)$ gravity is
\ben
S=\int d^4x\sqrt{-\G}\Big[\frac{f(\R,\T)}{16\pi G}+\La(X)\Big],
\label{13}
\een
where $f(\R,\T)$ is an arbitrary function of the Ricci scalar ($\R$) and trace of the energy-momentum tensor $(\T=\T^{\mu\nu}\G_{\mu\nu})$ and $L(X)$ is the non-canonical Lagrangian corresponding to the K-essence theory. Our revised action (\ref{13}) is clearly dependent on $\R$, $\T$, and $X(=\frac{1}{2}g^{\mu\nu}\nabla_{\mu}\phi\nabla_{\nu}\phi)$, rather than on the K-essence scalar field ($\phi$) explicitly. We can define the emergent energy-momentum tensor of this geometry as~\cite{Panda3}
\ben
\T_{\mu\nu}=-\frac{2}{\sqrt{-\G}}\frac{\partial\Big(\sqrt{-\G}\La(X)\Big)}{\partial \G_{\mu\nu}}
=\G_{\mu\nu}\La(X)-2\frac{\partial \La(X)}{\partial \G^{\mu\nu}},
\label{14}
\een
where $\big(-\G \big)^{1/2}=\big(-det({\G_{\mu\nu}})\big)^{1/2}$.

Using the variational principle, the field equation in this new geometry is written as~\cite{Panda3}
\ben
f_{\R}\R_{\mu\nu}+(\G_{\mu\nu}\bar{\square}-D_{\mu}D_{\nu})f_{\R}-\half \G_{\mu\nu}f(\R,\T)=8\pi G \T_{\mu\nu}-f_{\T}\T_{\mu\nu}-f_{\T}\Th_{\mu\nu}
\label{15}
\een
where $f_{\R}=\frac{\p f(\R,\T)}{\p \R},~f_{\T}=\frac{\p f(\R,\T)}{\p \T},~\bar{\square}=D_{\mu}D^{\mu}$ and $D_{\mu}$ is the covariant derivative with respect to the metric $\bar{G}_{\mu\nu}$.

And following \cite{Harko}, we have~\cite{Panda3}
\ben
\frac{\partial \T}{\partial \G^{\mu\nu}}=\frac{\partial(\T_{\alpha\beta}\G^{\alpha\beta})}{\partial \G^{\mu\nu}}=\T_{\mu\nu}+\bar{\Theta}_{\mu\nu},
\label{16}
\een
where
\ben
\bar{\Theta}_{\mu\nu}=\G^{\alpha\beta}\frac{\partial \T_{\alpha\beta}}{\partial\G^{\mu\nu}}.
\label{17}
\een

The condition for the preservation of the energy-momentum tensor ($D^{\mu}\T_{\mu\nu}=0$) is given by~\cite{Harko,Koivisto,Carvalho,Panda3,Panda2}
\ben
D^{\mu}\T_{\mu\nu}=\frac{f_{\T}}{8\pi G-f_{\T}}\Big[(\T_{\mu\nu}+\bar{\Theta}_{\mu\nu})D^{\mu}(\ln f_{\T})+D^{\mu}\bar{\Theta}_{\mu\nu}\Big]=0.
\label{18}
\een

Consider the flat FLRW metric as a background gravitational metric ($g_{\mu\nu}$) and then the corresponding K-essence emergent line element (\ref{8})  \cite{Panda3} can be written as ~\cite{Sharif3,Akbar,Kodama,Criscienzo2010,Faraoni} 
\ben
ds^2=h_{\a\b}dx^{\a}dx^{\b}+\tilde{r}^2d\Omega^2,
\label{19}
\een
where $\tilde{r}=a(t)r$; $x^0=t,~x^1=r$ with the two dimensional metric $h_{\a\b}=diag((1-\K),-\frac{a^2}{1-kr^2})$; $k$ is the cosmic curvature, and $d\Omega^2$ is the metric of 2-dimensional sphere with unit radius and $a(t)$ is the usual scale factor. 

By assuming the homogeneity of the K-essence scalar field, we choose it to be dependent on time only, denoted as $\phi=\phi(t)$~\cite{gm4,gm6,gm7,Panda3}.  Consequently, we get $\partial_{\rho}\phi\partial^{\rho}\phi=D_{\rho}\phi D^{\rho}\phi=\dot{\phi}^2$. It is crucial to make the choice as mentioned before, as the dynamical solutions of the K-essence scalar fields result in the spontaneous breaking of the Lorentz symmetry. Also, we should give careful consideration to the magnitude of $\dot{\phi}^2$ in Eq. (\ref{19}). It is clear that the condition $\dot\phi^{2}< 1$ must always be satisfied in order to obtain a meaningful signature of the emergent metric described by Eq.  (\ref{19})  \cite{Panda3}. In order to apply the K-essence theory, it is necessary for the condition $\dot\phi^{2}\neq 0$ to be satisfied. If we want to consider $\dot{\phi}^2$ as dark energy density $(\Omega_{DE})$ in units of the critical density, it is necessary for the value of $\dot{\phi}^2$ to be non-zero. It is always true that $\Omega_{Matter} +\Omega_{Radiation} +\Omega_{DE}= 1$. It is worth mentioning that the works cited in~\cite{gm1,gm2,gm3,gm6} provide evidence that $\dot{\phi}^2$ is associated with $(\Omega_{DE})$ and, as a result, it should fall within the range of $0<\dot{\phi}^2<1$.

By analyzing Eq. (\ref{6}), we can establish a connection between the Hubble parameter ($H(t)$) and the K-essence scalar field as \cite{Panda2, Panda3}
\ben
3\frac{\dot a}{a}=3H(t)=-\frac{\ddot\phi}{\dot\phi(1-\dot\phi^{2})},
\label{20}
\een
with the fact that $\dot{a}\neq 0$. Assuming that the energy-momentum tensor $(\bar{T}_{\mu\nu})$ takes the form of an ideal fluid, we can express it as
\ben
\T_{\mu}^{\nu}&=& diag(\bar{\rho},-\bar{p},-\bar{p},-\bar{p})=(\bar{\rho} +\bar{p})u_{\mu}u^{\nu}-\delta_{\mu}^{\nu} \bar{p}\nonumber\\
\T_{\mu\nu}&=&\G_{\mu\a}\T^{\a}_{\nu},
\label{21}
\een
where $\bar{p}$ is pressure and $\bar{\rho}$ is the energy density of the cosmic fluid in K-essence emergent geometry. In the co-moving frame, the values of $u^{0}$ and $u^{\a}$ are $1$ and $0$ respectively, where $\a$ ranges from 1 to 3 in the K-essence emergent gravity spacetime. The Lagrangian form in Eq. (\ref{7}) indicates that the perfect fluid model with zero vorticity is applicable in K-essence theory~\cite{Vikman,Babichev2,Panda3}. Additionally, the pressure can be expressed solely in terms of the energy density.

By considering the $(00)$ and $(11)$ components of the field Eq. (\ref{15}), we can express the Friedmann equations as follows:
\ben
3\Big[H^2+\frac{k(1-\K)}{a^2}\Big]&&=\frac{1}{f_{\R}}\Big[\frac{1-\K}{2}(f(\R,\T)-\R f_{\R})-3Hf_{\R\R}\dot{\R}-f_{\T}\La(1-\K)+f_{\T}\frac{\dot{\f}^4(1+2\K)(5-2\K)}{(1-\K)^{1/2}}\Big]\nonumber\\
&&+8\pi G_{eff}\rh(1-\K)
\label{22}
\een
and
\ben
&&-\Big[2\dot{H}+3H^2(1-2\K)+\frac{k(1-\K)}{a^2}\Big]=\frac{1}{f_{\R}}\Big[f_{\R\R\R}\dot{\R}^2+f_{\R\R}\ddot{\R}+2Hf_{\R\R}\dot{\R}(1-\K)\Big]\nonumber\\
&&-\frac{1-\K}{f_{\R}}\half (f-\R f_{\R})-8\pi G_{eff} \bar{p}(1-\K)+\frac{1-\K}{f_{\R}}f_{\T}\La,
\label{23}
\een
with
\ben 
G_{eff}=\frac{1}{f_{\R}(\R,\T)}\Big(G+\frac{f_{T}(\R,\T)}{8\pi}\Big),
\label{24}
\een 
where $G$ is the universal constant and for homogeneous K-essence scalar field ($\phi(t,x)\equiv\phi(t)$), $\La(X)$ (\ref{7}) can be expressed as $\La(X)\equiv\La=\Big(1-\sqrt{1-\K}\Big)$ for FLRW background gravitational metric. Equations (\ref{22}) and (\ref{23}) can also be recast as
\ben
3H^2+\frac{3k(1-\K)}{a^2}=8\pi G_{eff}(1-\K)(\rh+\rh_{d})
\label{25}
\een
and 
\ben
-2\dot{H}+\frac{2k}{a^2}(1-\K)(1-3\K)=8\pi G_{eff}(1-\K)\Big[(\rh+\rh_{d})(1-2\K)-\bar{p}+\bar{p}_{d}\Big]
\label{26}
\een 
where the energy density and pressure contributions, denoted as $\rh_{d}$ and $\bar{p}_{d}$, respectively, are associated with the non-canonical K-essence $f(\R,\T)$ theory, which is referred to as the dark components of the universe.

These components can be expressed as: 
\ben
\rh_{d}=\frac{1}{8\pi G\F(1-\K)}\Big[\frac{1-\K}{2}\Big(f(\R,\T)-\R f_{\R}\Big)-3Hf_{\R\R}\dot{\R}-f_{\T}\La\Big(1-\K\Big)+f_{\T}\frac{\dot{\f}^4(1+2\K)(5-2\K)}{(1-\K)^{1/2}}\Big]
\label{27}
\een
and 
\ben
\bar{p}_{d}=\frac{1}{8\pi G\mathcal{F}(1-\K)}\Big[f_{\R\R\R}\dot{\R}^2+f_{\R\R}\ddot{\R}+2Hf_{\R\R}\dot{\R}(1-\K)-\half (1-\K)(f-\R f_{\R})-(1-\K)f_{\T}\La\Big]
\label{28}
\een 

where $\F=1+\frac{f_{\T}}{8\pi G}$. In our theory, it is crucial to note that the pressure of the K-essence field ($\bar{p}$) in Eqs. (\ref{23}) and (\ref{26}) cannot be assumed to be zero. In K-essence geometry, the scalar fields are minimally coupled with the usual gravity via the K-essence Lagrangian, which acts as the pressure~\cite{Picon2}. Thus, in contrast to the findings of ~\cite{Sharif3}, we are unable to include the pressureless dust regime in our theory.

One can calculate the Equation of State (EoS) parameter for the dark components as follows:

\begin{equation}
\bar{\omega}_{d}=-1+\frac{f_{\R\R\R}\dot{\R}^2+f_{\R\R}\ddot{\R}-Hf_{\R\R}\dot{\R}-f_{\T}\frac{\dot{\f}^4(1+2\K)(5-2\K)}{2(1-\K)^{1/2}}}{\frac{1-\K}{2}(f-\R f_{\R})-3Hf_{\R\R}\dot{\R}-f_{\T}\La\Big(1-\K\Big)+f_{\T}\frac{\dot{\f}^4(1+2\K)(5-2\K)}{2(1-\K)^{1/2}}}.\\
\label{29}
\end{equation}

The extra terms that arise in usual $f(R,T)$ gravity are considered as the dark components of the universe \cite{Sharif3}. The dark components usually can be thought of as the combination of dark matter and dark energy of the universe. In our case the dark components Eqs. (\ref{27}), (\ref{28}), (\ref{29}) have been modified due to the presence of the $\dot{\phi}^2$ term both intensively and extensively.

The semi-conservation equation for the coupling system, which consists of gravity coupled with a K-essence scalar field, may be expressed as:
\ben
\dot{\rh}_{tot}+3H\Big[\rh_{tot}(1-2\K)+\bar{p}_{tot}\Big]=\bar{q}_{tot}
\label{30}
\een
where
\ben
\bar{q}_{tot}=\frac{3}{8\pi G}\Bigg[\frac{\partial}{\partial t}\Big(\frac{f_{\R}}{\mathcal{F}}\Big)\Big(\frac{H^2}{1-\K}+\frac{k}{a^2}\Big)+\frac{f_{\R}}{\mathcal{F}}\Big[-\frac{6H^3\K}{1-\K}-\frac{6kH\K}{a^2}\Big]\Bigg]
\label{31}
\een

The findings of our investigation Eqs. (\ref{25})--(\ref{31}) vary significantly from those derived by Sharif et. al.~\cite{Sharif3} in their article. The discrepancies emerge from the coupling geometry of our K-essence type, where the combination of gravity and the scalar field creates a new kind of geometry, (\ref{5}) or (\ref{8}). By setting $\dot{\phi}^{2}=0$ and hence $\La(X)=\La_m$, we revert to the conventional outcomes of the standard $f(R,T)$ gravity as presented in the work of Sharif et al.~\cite{Sharif3}.

\section{Thermodynamics of $f(\R,\T)$ gravity}

This section investigates the validity of the first and second laws of thermodynamics in the context of $f(\R,\T)$ gravity for the FLRW background metric.

\subsection{First Law of Thermodynamics}

In this subsection, we investigate the reliability of the first law of thermodynamics in the context of $f(\R,\T)$ gravity at the apparent horizon of the K-essence FLRW spacetime (\ref{19}). The definition of ``marginally trapped surface with vanishing expansion" in the context of the FLRW universe refers to the apparent horizon of the universe~\cite{Cai,Cai3,Sharif3}.  This is a dynamic surface that changes as the cosmos evolves. The apparent horizon is a surface where the rate of expansion of outgoing light rays (null rays) is zero. The mathematical expression for this is that the condition at the apparent horizon is given by $h^{\a\b}\partial_{\a}\r\partial_{\b}\r=0$. The equation provides the value of the radius ($\r_{A}$) of the apparent horizon, which can be expressed as~\cite{Cai,Sharif3}
\ben
\tilde{r}_A=\sqrt{\frac{1-\K}{H^2+\frac{k}{a^2}(1-\K)}}
\label{32}
\een
and the corresponding surface gravity ($\kappa_{sg}$) can be evaluated as~\cite{Kodama,Criscienzo2010,Faraoni,Sharif3}
\ben
\kappa_{sg}&&=\frac{1}{2\sqrt{-h}}\partial_{\a}(\sqrt{-h}h^{\a\b}\partial_{\b}\tilde{r}_{A})\nonumber\\
&&=\frac{1}{1-\K}\frac{\tilde{r}}{2}(\dot{H}+2H^2)-\frac{3}{2a^2}\frac{\K}{1-\K}H\sqrt{a^2-k\tilde{r}^2}-\frac{k\tilde{r}}{2a^2}
\label{33}
\een
where the associated temperature at the apparent horizon is $T_{h}=\frac{|\kappa_{sg}|}{2\pi}$. It is important to note that in black hole thermodynamics, surface gravity typically plays a role analogous to temperature. However, in fully dynamic situations, surface gravity does not correspond directly to the temperature of a thermal spectrum. Despite this, it is still expected to significantly influence Hawking-like radiation, even in non-equilibrium processes. Surface gravity is traditionally defined on a Killing horizon, which is effective for stationary black holes, but this method fails in dynamic scenarios where no Killing horizon exists. To define surface gravity for a dynamical metric, the Kodama vector field is employed. In the context of spherically symmetric dynamical spacetimes, this field allows for the definition of a conserved current and energy in the absence of a global timelike Killing vector \cite{Sharif3,Akbar,Kodama,Criscienzo2010,Faraoni}. Thus, the Kodama vector can be seen as a generalization of the Killing vector field, applicable to spacetimes without a Killing vector, and has been used in the thermodynamics of dynamic spacetimes. For calculating surface gravity (\ref{33}), we rely on this Kodama vector concept.

It is evident that the radius of the apparent horizon $\r_{A}$ (\ref{32}), and the associated surface gravity $\kappa_{sg}$ (\ref{33}), deviate from the usual $f(R,T)$ gravity~\cite{Sharif3} because of the inclusion of the kinetic component of the K-essence scalar field ($\dot\phi^{2}$). It should be noted that in the study conducted by Sharif et. al. \cite{Sharif3}, the apparent horizon and surface gravity are exclusively determined by the scale factor ($a$) and the Hubble parameter ($H$) and its derivative. In our case, the scale factor as well as the Hubble parameter is influenced by the scalar field ($\K$). In this emergent geometry, some extra term arises due to the coupling of matter and geometry. Here, we graphically analyze the illustration of $\kappa_{sg}$ for two separate scale parameter selections and different curvature values ($k=-1,0,1$), which have some physical relevance.\\

{\bf Case I:} Firstly, we consider the exponential scale parameter expressed as
\ben
a(t)=e^{H_{0}t},
\label{34}
\een
where $H_{0}$ is a constant value of the Hubble parameter bearing different values at different epochs. Note that this type of scale factor can be configured with late time acceleration ~\cite{Weinberg} as well as inflation~\cite{Peebles, Liddle, Mukhanov1} scenarios.\\

{\bf Case II:} Secondly, we consider the power law scale factor~\cite{Kadam,Bahamonde,Ye} expressed as
\ben
a(t)=C \Big(\frac{t}{t_{0}}\Big)^m,~m>0,~C=\text{constant},
\label{35}
\een
From Eq. (\ref{20}), $\K$ can be expressed as 
\ben
\K=\frac{1}{1-a^6(t)}.
\label{36}
\een

\begin{figure*}[h]
\begin{minipage}[b]{0.4\linewidth}
\centering
 \begin{subfigure}[b]{0.9\textwidth}
    \includegraphics[width=6cm]{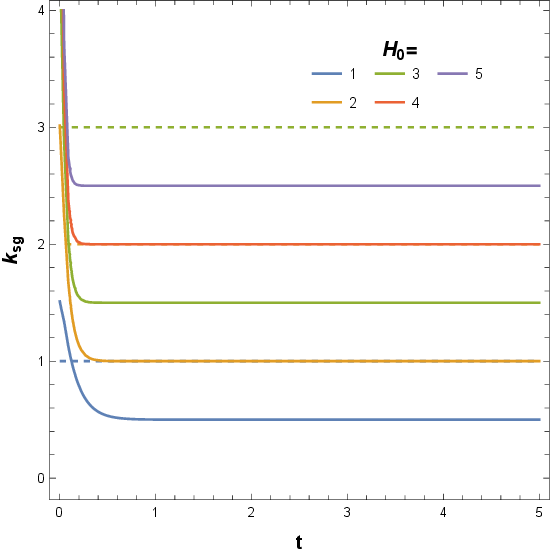}
       \caption{Comparative study of $\k_{sg}$ with time for usual case (dotted line) and modified case (solid line) for $k=0$ using exponential scale factor $H_{0}=1,2,3,4,5$}
        \label{Fig1a}
    \end{subfigure}
\end{minipage}
\hspace{2cm}
\begin{minipage}[b]{0.4\linewidth}
\centering
 \begin{subfigure}[b]{0.9\textwidth}
    \includegraphics[width=6cm]{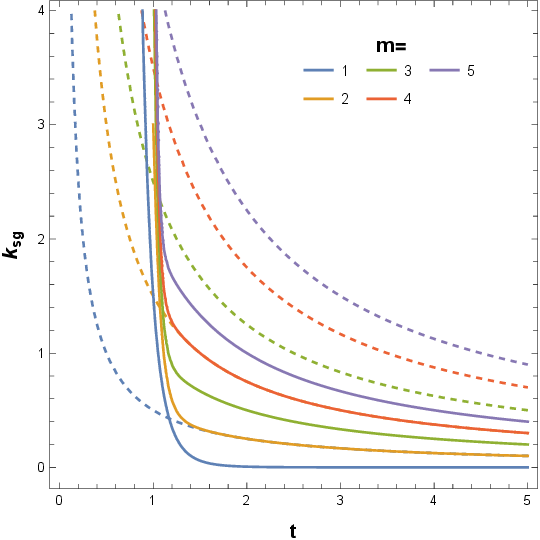}
        \caption{ 
 Comparative study of $\k_{sg}$ with time for usual case (dotted line) and modified case (solid line) for $k=0$ using power law scale factor $m=1,2,3,4,5$}
        \label{Fig1b}
    \end{subfigure}
\end{minipage}
\caption{Variation $\k_{sg}$ with time for $k=0$}
\label{Fig1}
\end{figure*}

\begin{figure*}[h]
\begin{minipage}[b]{0.4\linewidth}
\centering
 \begin{subfigure}[b]{0.8\textwidth}
    \includegraphics[width=6cm]{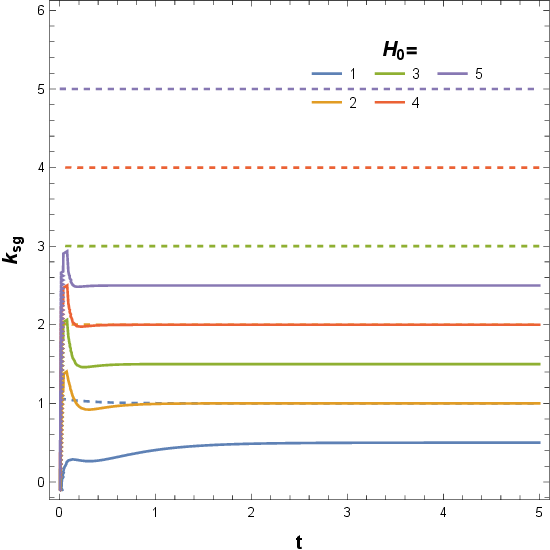}
       \caption{ Comparative study of $\k_{sg}$ with time for usual case (dotted line) and modified case (solid line) for $k=1$ using exponential scale factor $H_{0}=1,2,3,4,5$}
        \label{Fig2a}
    \end{subfigure}
\end{minipage}
\hspace{2cm}
\begin{minipage}[b]{0.4\linewidth}
\centering
 \begin{subfigure}[b]{0.8\textwidth}
    \includegraphics[width=6cm]{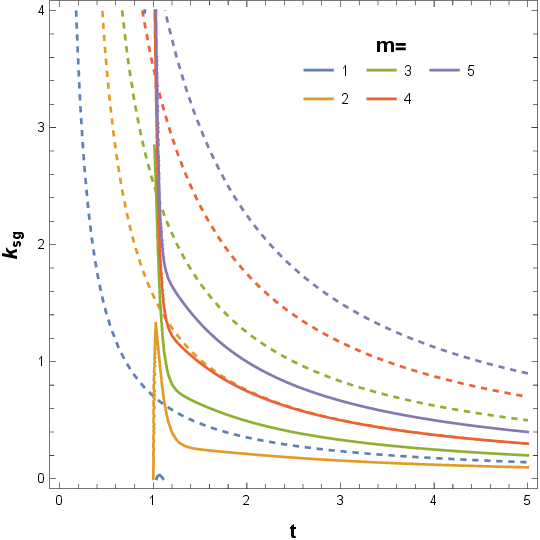}
        \caption{ 
 Comparative study of $\k_{sg}$ with time for usual case (dotted line) and modified case (solid line) for $k=1$ using power law scale factor $m=1,2,3,4,5$}
        \label{Fig2b}
    \end{subfigure}
\end{minipage}
    \caption{Variation $\k_{sg}$ with time for $k=1$}
\label{Fig2}
\end{figure*}

Now, we graphically represent the surface gravity at the apparent horizon for the two cases mentioned above. In the graphs (Figs. (\ref{Fig1}) -- (\ref{Fig3})) below the dotted lines represent the characteristics in the usual case as described by Sharif et al.~\cite{Sharif3}, while the solid line represents the characteristics in our modified case.

Figure 1(a) illustrates the comparison of the time-dependent behavior of the $\k_{sg}$ parameter between the standard $f(R,T)$ gravity and a non-canonical (K-essence) $f(\R,\T)$ gravity scenario, both with $k=0$, using the exponential scale factor (\ref{34}). Typically, surface gravity diminishes with time, beginning from a larger magnitude and eventually reaching a constant value. In our scenario, the parameter $\k_{sg}$ exhibits comparable fluctuations i.e., at the initial time ($t\approx 0$), the value of $\k_{sg}$ undergoes a rapid change.

On the other hand, Fig. 1(b) represents the $\k_{sg}$ {\it vs.} time plot using the power law scaling factor (\ref{35}). In addition to the initial value, there is also a disparity in stiffness over the early period. Typically, the decrease of $\k_{sg}$ is less in the unmodified case compared to the modified case. The values of $\k_{sg}$ for each $m$ value are lower than in the ordinary scenario. The initial conditions of the two situations also vary. The value of $\k_{sg}$ is missing at an early period ($t\simeq 0$) for the modified instance, yet the graphs have nearly the same characteristics. In this figure, we can see that at the time $t\approx 1$, there is a sudden decrease in surface gravity, like a step function for the modified case. The surface gravity decreases abruptly at the apparent horizon due to the expansion rate of the cosmos, which may cause distances to rise at a rate that opposes the local gravitational force. This event might have happened because the usual gravitational spacetime and the scalar (matter) field are coupled in our scenario. This may cause a sudden change in the surface gravity around the time $t\simeq 1$.

Near $t=0$, the variable $\k_{sg}$ has a finite positive value in typical scenarios. However, in our specific case, no finite value has been attained for $\k_{sg}$ in the two figures mentioned above. If we consider $t\simeq 0$ as the starting point of the universe (Big Bang), it is not possible to assign a definite value to it at that moment. At that time, we cannot determine the exact temperature associated with surface gravity, possibly due to significant fluctuations. However, the study of temperature fluctuation is beyond the scope of this article. In a different context, Gangopadhyay~\cite{Gangopadhyay2010} evaluates the temperature fluctuations in the early universe using the K-essence model.

Figure 2(a) represents a comparison of the surface gravity $(\k_{sg})$ for positive curvature and exponential scale factor (Eq. (\ref{34})). In the usual case, the value of $\k_{sg}$ has a constant behavior. In the modified case, the value of $\k_{sg}$ starts after $t=0$, and then it shows some fluctuation as similar in the case of Fig. 1(a), then it comes to a constant value. Figure 2(b) represents the same for the power law scale factor (Eq. ({\ref{35})). However, the nature of the usual and modified cases remains the same in this case but as before the stiffness varies. Also, the discrepancy persists at an early time. The sudden drop in $\k_{sg}$ exists in this case too resulting in a step function-like behavior in Fig. 2(b).

\begin{figure*}[h]
\begin{minipage}[b]{0.4\linewidth}
\centering
 \begin{subfigure}[b]{0.8\textwidth}
    \includegraphics[width=6cm]{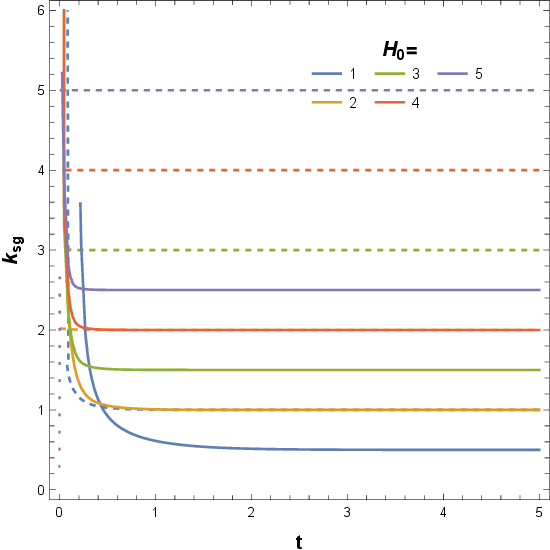}
       \caption{ Comparative study of $\k_{sg}$ with time for usual case (dotted line) and modified case (solid line) for $k=-1$ using exponential scale factor $H_{0}=1,2,3,4,5$}
        \label{Fig3a}
    \end{subfigure}
\end{minipage}
\hspace{2cm}
\begin{minipage}[b]{0.4\linewidth}
\centering
 \begin{subfigure}[b]{0.8\textwidth}
    \includegraphics[width=6cm]{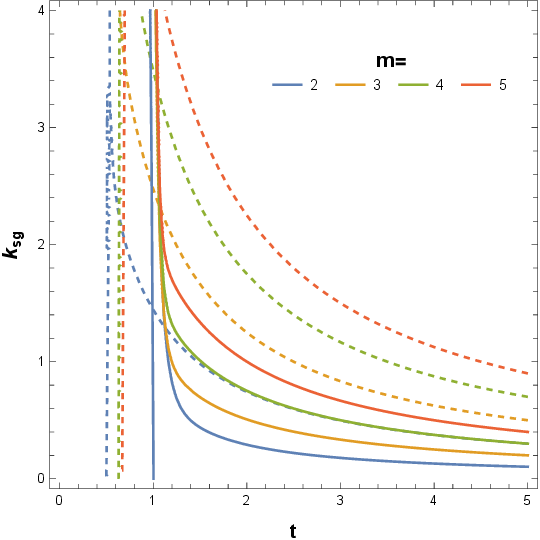}
        \caption{ 
 Comparative study of $\k_{sg}$ with time for usual case (dotted line) and modified case (solid line) for $k=-1$ using power law scale factor $m=1,2,3,4,5$}
        \label{Fig3b}
    \end{subfigure}
\end{minipage}
\caption{Variation $\k_{sg}$ with time for $k=-1$}
\label{Fig3}
\end{figure*}

Figure 3(a) represents the comparative study of $\k_{sg}$ for negative curvature ($k=-1$) for exponential scale factor (Eq. (\ref{34})). For this case the behaviour of the two graphs is similar. There is a shift in the time scale only.  Fig. \ref{Fig3b} represents the same for the power law scale factor (Eq. (\ref{35})). The value of $\k_{sg}$ of the modified case is lower than the usual case in both the Figs. 3(a) and 3(b). The unusual behavior at the early time shown in our case may be the indication of fluctuation due to the scalar field~\cite{Mukhanov1}. Due to the presence of this fluctuation in the physical parameter, it may arise in scalar field cosmology.

The Bekenstein-Hawking relation~\cite{Bardeen,Hawking,Bekenstein} $S_{h} = A/4G$ in GR gives the horizon entropy, where $A=4\pi \r_{A}^2$ is the apparent horizon's area. Wald~\cite{Wald} proposed that the entropy of black hole solutions with bifurcate Killing horizons may be understood as a Noether charge entropy within the framework of modified gravity theories. It is determined by how the modified gravitational theories' Lagrangian density evolves with respect to the Riemann tensor. In terms of effective gravitational coupling, a quarter of the horizon area is equivalent to Wald entropy, i.e., $S_{h}=A/4G_{eff}$~\cite{Hayward1,Hayward2}. Following~\cite{Sharif3}, in our $f(\R,\T)$ gravity the Wald entropy can be expressed as 
\ben
S_{h}=\frac{Af_{\R}}{4G\F}=\frac{Af_{\R}}{4G+\frac{f_{\T}}{2\pi}}.
\label{37}
\een

This proposition is not just a correspondence of the Bekenstein-Hawking relation, rather, it has been extensively derived for modified gravity in \cite{Brustein}.

Taking the derivative of Eq. (\ref{32}) and using (\ref{37}) we get
\ben
d\r_{A}=\frac{H\r_{A}^3}{1-\K}4\pi G\frac{\mathcal{F}}{f_{\R}}\Big[(1-\K)(1-2\K)\rh_{tot}+(1-\K)\bar{p}_{tot}\Big]dt+3H\r_{A}\K dt,
\label{38}
\een 
where $\rh_{tot}=\rh+\rh_{d}$ and $\bar{p}_{tot}=-\bar{p}+\bar{p}_{d}$. 

The small variation in the apparent horizon's radius during a given time interval $dt$ is denoted as $d\r_{A}$. Using (\ref{37}) and (\ref{38}) we can write 
\ben
\frac{dS_{h}}{2\pi \r_{A}}=\frac{4\pi\r_{A}^3}{1-\K}\Bigg[\Big[(1-\K)(1-2\K)\rh_{tot}+(1-\K)\bar{p}_{tot}+\frac{3\K f_{\R}}{4\pi G\r_{A}^2\F}\Big]Hdt+\frac{\r_{A}f_{\R}}{2G}d\Big(\frac{1}{\F}\Big)+\frac{\r_{A}}{2G\F}df_{\R}\Bigg].
\label{39}
\een
 
It follows that if we multiply the two sides of this equation by a factor ($1-\frac{\dot{\r}_A}{2H\r_{A}}$), we have 
\ben
T_{h}dS_{h}&&=\frac{4\pi \r_{A}^3}{1-\K}\Big[(1-\K)(1-2\K)\rh_{tot}+(1-\K)\bar{p}_{tot}+\frac{3\K}{f_{\R}}{4\pi}G\r_{A}^2\F\Big]Hdt\nonumber\\
&&-\frac{2\pi \r_{A}^2}{1-\K}\Big[(1-\K)(1-2\K)\rh_{tot}+(1-\K)\bar{p}_{tot}+\frac{3\K}{f_{\R}}{4\pi}G\r_{A}^2\F\Big]d\r_{A}+\frac{r_{A}^2f_{\R}\pi}{G}T_{h}d(\frac{1}{\F})+\frac{\pi\r_{A}^2T_{h}}{G\F}df_{\R}.
\label{40}
\een
 
We now characterize cosmic energy inside the apparent horizon. The total matter-energy within a sphere is represented by the Misner-Sharp energy~\cite{Misner,Bak,Kodama,Criscienzo2010,Faraoni} in the FLRW universe, a model of a homogeneous and isotropic expanding or contracting universe. This energy might be thought of as the edge of the finite area under consideration. When the energy is projected into the universe's apparent horizon, it obeys the unified first law of thermodynamics. In a similar manner as~\cite{Sharif3}, the Misner-Sharp energy~\cite{Misner,Bak,Gong,Wu} can be defined in $f(\R,\T)$ gravity as 
\ben
E=\frac{\r_{A}}{2G_{eff}}.
\label{41}
\een

Considering the volume element $V=\frac{4}{3}\pi \r_{A}^3$ and using Eqs. (\ref{25}) and (\ref{32}) the energy (\ref{41}) can be re-written as
\ben
E=\frac{3V}{4\pi \r_{A}^2}\frac{1}{2G_{eff}}=V\rh_{tot}.
\label{42}
\een

It stands for the total energy contained within the sphere of radius $\r_{A}$. Just like in the case of regular $f(R,T)$ gravity~\cite{Sharif3}, in order for the energy to be positive ($E>0$), it is necessary for the effective coupling constant $G_{eff}=\frac{G\F}{f_{\R}}$ to be positive, which is also true in this case. The only difference in this case is the term $f_{\R}$ is defined in the context of the K-essence $f(\R,\T)$ geometry.

From (\ref{26}) and (\ref{42}) it can be written 
\ben
dE&&=4\pi\r_{A}^2\rh_{tot}d\r_{A}+\frac{\r_{A}(1-\K)}{2G\F}df_{\R}+\frac{\r_{A}(1-\K)f_{\R}}{2G}d(\frac{1}{\F})\nonumber\\
&&-\Big[4\pi\r_{A}^3H\Big((1-2\K)\rh_{tot}+\bar{p}_{tot}\Big)+\frac{\r_{A}^3}{2G_{eff}}\Big(\frac{6k}{a^2}\K(1-\K)-6H^2\K\Big)\Big]dt
\label{43}
\een
 
Using (\ref{43}) in (\ref{40}) we can write 
\ben
&&(1-\K)T_{h}dS_{h}=-dE'+W'dV+\frac{\r_{A}}{2G\F}(1-\K)\Big[1+2\pi \r_{A}T_{h}\Big]df_{\R}+\frac{\r_{A}(1-\K)f_{\R}}{2G}\Big[1+2\pi \r_{A}T_{h}\Big]d\Big(\frac{1}{\F}\Big),
\label{44}
\een
where
\ben
dE'=dE-\K\Big[4\pi H\r_{A}^3\Big((1-2\K)\rh_{tot}-\bar{p}_{tot}\Big)+\frac{3\r_{A}^2 H}{G_{eff}}+\frac{\r_{A}^3}{3G_{eff}}\Big(\frac{k}{a^2}(1-\K)-H^2\Big)\Big]dt
\label{45}
\een
and work density~\cite{Hayward1,Sharif3}
\ben
W'=W+\K\Big[2\pi \r_{A}^2(3\rh_{tot}+\bar{p}_{tot})+\frac{3}{G_{eff}}\Big]d\r_{A}
\label{46}
\een
with $W=\frac{1}{2}(\rh_{tot}-\bar{p}_{tot})$.

Equation (\ref{44}) can be expressed in a compact form
\ben
(1-\K)T_{h}dS_{h}+(1-\K)T_{h}d_{j}S_{h}=-dE'+W'dV
\label{47}
\een
where
\ben
d_{j}S_{h}&&=-\frac{\r_{A}}{2GT_{h}}\Big[1+2\pi\r_{A}T_{h}\Big]d\Big(\frac{f_{\R}}{\F}\Big)\nonumber\\
&&=-\frac{\F(E+S_{h}T_{h})}{T_{h}f_{\R}}d\Big(\frac{f_{\R}}{\F}\Big).
\label{48}
\een

On comparison with GR, Gauss-Bonnet gravity, and Lovelock gravity~\cite{Cai,Akbar} the usual $f(R,T)$ gravity produces the extra entropy production term  $d_{j}S_{h}$ \cite{Sharif3} due to the coupling, which has been modified in our geometry through the modified $f_{\R}$ and $f_{\T}$ terms.

\subsection{Generalized second law of thermodynamics in $f(\R,\T)$ gravity}

According to~\cite{Sharif3,Davies1987,Wu} the generalized second law of thermodynamics (GSLT) can be expressed as
\ben
\dot{S}_{h}+d_{j}\dot{S}_{h}+\dot{S}_{tot}\geq 0,
\label{49}
\een
where $S_{h}$ is the horizon entropy, $d_{j}\dot{S}_{h}=\partial_{t}(d_{j}S_{h})$ and $S_{tot}$ is the total entropy resulting from all the matter and energy sources inside the horizon in the modified $f(\R,\T)$ gravity. In~\cite{Wu,Izquierdo}, it is noted that ordinary matter can be seen as a representation of a mixture of fields. These fields can be in a pure state or not and they have their own entropy. With the inclusion of other matter fields and energy components (e.g. interactive fields or energy), the total temperature $T_{tot}$ along with the total entropy $(S_{tot})$ gives
\ben
T_{tot}dS_{tot}=d(\rh_{tot}V)+\bar{p}_{tot}dV,
\label{50}
\een
which is Gibb's equation for the fluid of matter (scalar fields) and energy~\cite{Izquierdo}. Within the horizon, $T_{tot}$ represents the temperature as a whole. Here, we expressed the above Gibb's equation in the context of K-essence $f(\R,\T)$ gravity such that when we withdraw the coupling factor, i.e., if we consider non-interaction between gravity and scalar fields, we get back to the original Gibb's equation stated in~\cite{Sharif3,Wu,Izquierdo}.  The temperature of the apparent horizon is assumed to be proportional to $T_{tot}$~\cite{Wu,Bamba} i.e., $T_{tot}=bT_{h}$, where $0<b<1$, to guarantee that the temperature is both below the horizon and positive.
Using (\ref{45}) and (\ref{50}) in (\ref{49}) we get
\ben
\dot{S}_{h}+d_{j}\dot{S}_{h}+\dot{S}_{tot}=\frac{24\pi \Xi}{\r_{A}b\mathcal{R}}\geq 0
\label{51}
\een
where
\ben
\Xi&&=\frac{\rh_{tot}\dot{V}}{1-\K}\Big[1-\frac{b}{2}+\K(b-1)\Big]+\frac{V\dot{\rh}_{tot}}{(1-\K)}(1-b+\K)+\frac{\bar{p}_{tot}\dot{V}}{(1-\K)}(1-\frac{b}{2}-\K)\nonumber\\
&&+\frac{4\pi \r_{A}^3 b \K}{(1-\K)}\Big[2\pi\r_{A}(3\rh_{tot}+\bar{p}_{tot})+\frac{1}{G_{eff}}\Big]\dot{\r}_{A}
\label{52}
\een
and 
\ben
\mathcal{R}=\R-\frac{18\K H}{(1-\K)}\Big(H+\frac{\sqrt{a^2-k\r_{A}^2}}{a^2\r_{A}}\Big)+\frac{12k}{a^2}.
\label{53}
\een

The additional interaction terms due to the presence of $\K$ modify the GSLT (\ref{51}) in our case. If we consider $\K=0$ we would eventually get back the usual GSLT in the usual $f(R,T)$ gravity ~\cite{Sharif3}. We expressed the terms in Eqs. (\ref{51}) and (\ref{52}) in a manner that makes the GSLT for our instance look like the structure suggested in~\cite{Sharif3}. However, the underlying geometry greatly differs between the two situations, which should result in additional characteristics beyond the ordinary case.

\section{Study of Entropy}

In this part, we want to demonstrate the validity of GSLT (\ref{49}) or (\ref{51}).

Let us denote the rate of change of entropy for horizon entropy ($\dot{S}_h$), entropy from inside the horizon ($\dot{S}_{tot}$) and extra entropy coming from interaction ($d_j\dot{S}_h$) as a sum
\ben
\dot{S}=\dot{S}_h+d_j\dot{S}_h+\dot{S}_{tot}.
\label{54}
\een

According to the generalized second law of thermodynamics (GSLT) this term should be greater than zero~\cite{Wu,Sharif3}. To check this in our context we calculate the terms $\Xi,~\mathcal{R}$ in Eq.s (\ref{52}) and (\ref{53}) respectively. Using (\ref{25}) and (\ref{26}) we can find out $\rh_{tot},~\dot{\rh}_{tot}~\text{and}~\bar{p}_{tot}$. $\dot{V}$ can be calculated as $4\pi\r_{A}^2\dot{\r}_{A}$. The value of $\r_{A}$ and $\dot{\r}_{A}$ can be formulated from Eq. (\ref{32}). Now we have to choose a particular function of $f(\R,\T)$. We decide to take it as Starobinsky type function~\cite{Moraes,Star,Panda3} which says
\ben
f(\R,\T)=\R+\a\R^2+\lambda \T
\label{55}
\een
where $\a$ and $\lambda$ are constants. 

Also, consider that the scalar field behaves as~\cite{gm4,Panda3}
\ben
\K=e^{-t/\tau}
\label{56}
\een
where $\tau$ is a positive constant, maintaining the conditions imposed on $\K$ as $0<\K<1$. 

The relationship between the scale factor ($a(t)$) and the scalar filed ($\phi$) in Eq. (\ref{20}) has also been used to derive 
\ben
a(t)=\Big(\frac{1-\K}{\K}\Big)^{1/6}.
\label{57}
\een

\subsection{Variation of $\dot{S}$}

Using Eqs. (\ref{55}), (\ref{56}) and (\ref{57}), we derive from (\ref{51})
\ben
\dot{S}=\frac{N}{D},
\label{58}
\een
where the expression of the numerator (N) and the denominator (D) has been expressed in (\ref{A1}) and (\ref{A2}) in {\bf Appendix A} (Sec-\ref{Appendix-A}).

Now let's plot this function ($\dot{S}$) with respect to the cosmic time ($t$) for some particular choices of parameters ($\alpha,~\lambda,~\tau,~\text{and}~b$). We take the positive ($k=1$), zero ($k=0$) and negative curvature ($k=-1$) values in each graph. The values of $\a$ and $\lambda$ are set to $1$ for simplicity.

\begin{figure*}[h]
\hspace{1cm}
\begin{minipage}[b]{0.4\linewidth}
\centering
 \begin{subfigure}[b]{1.0\textwidth}
        \includegraphics[width=\textwidth]{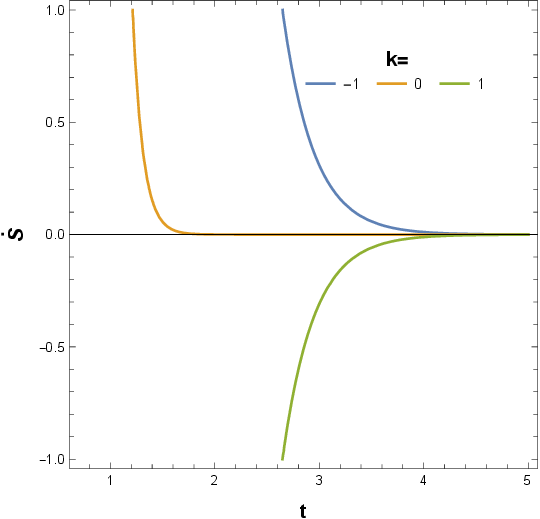}
        \caption{Variation of $\dot{S}$ with $t$ for $\alpha=1,~\lambda=1,~\tau=0.1,~b=0.1$}
        \label{Fig.4a}
    \end{subfigure}
\end{minipage}
\hspace{2cm}
\begin{minipage}[b]{0.4\linewidth}
\centering
 \begin{subfigure}[b]{1.0\textwidth}
        \includegraphics[width=\textwidth]{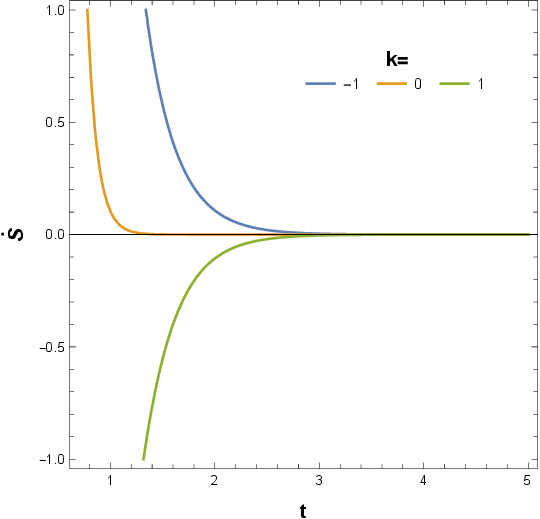}
        \caption{Variation of $\dot{S}$ with $t$ for $\alpha=1,~\lambda=1,~\tau=0.1,~b=0.9$}
        \label{Fig.4b}
    \end{subfigure}
\end{minipage}
\caption{Variation of $\dot{S}$ with $t$}
\label{Fig.4}
\end{figure*}

\begin{figure*}[h]
\hspace{1cm}
\begin{minipage}[b]{0.4\linewidth}
\centering
 \begin{subfigure}[b]{1.0\textwidth}
        \includegraphics[width=\textwidth]{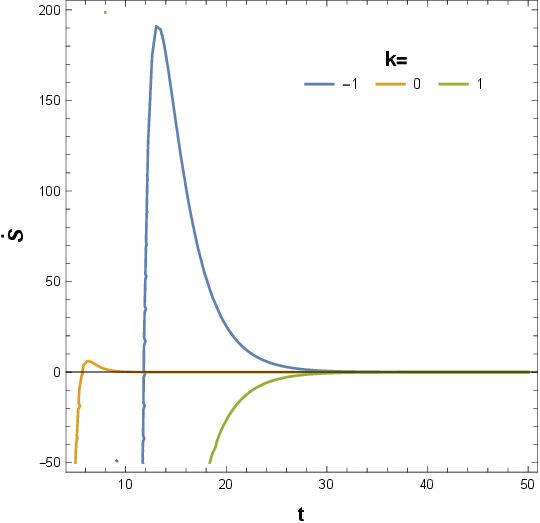}
        \caption{Variation of $\dot{S}$ with $t$ for $\alpha=1,~\lambda=1,~\tau=0.9,~b=0.1$}
        \label{Fig.5a}
    \end{subfigure}
\end{minipage}
\hspace{2cm}
\begin{minipage}[b]{0.4\linewidth}
\centering
 \begin{subfigure}[b]{1.0\textwidth}
        \includegraphics[width=\textwidth]{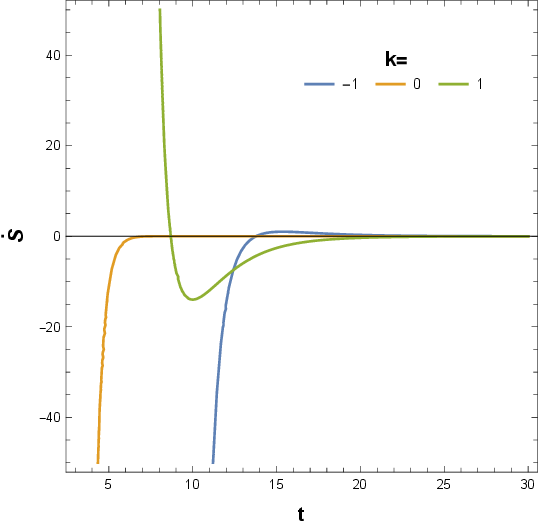}
        \caption{Variation of $\dot{S}$ with $t$ for $\alpha=1,~\lambda=1,~\tau=0.9,~b=0.9$}
        \label{Fig.5b}
    \end{subfigure}
\end{minipage}
\caption{Variation of $\dot{S}$ with $t$}
\label{Fig.5}
\end{figure*}

Figure 4(a) represents the variation of $\dot{S}$ with time ($t$) for $\tau=0.1$ and $b=0.9$. This graph shows negative behavior at first, then it becomes constant at $t=0$ for $k=1$. For $k=0~\&~-1$, $\dot{S}$ is always positive. Fig. (\ref{Fig.4b}) which has been plotted for $(\dot{S})$ with $\a=1,~\lambda=1,~\tau=0.1~\&~b=0.9$ shows the similar type of behavior. For all the cases of $k=-1,0,1$, $\dot{S}$ becomes constant at zero after a time interval, indicating the universe's constant entropy. In comparison with Figs. 4(a) and 4(b) we get to see that when $b$ is greater, the entropy becomes constant at an earlier stage of time. From these two graphs, it may be said that the universe with positive curvature seems irrelevant in the context of thermodynamics. The greater value of $b~(\to 1)$ means the horizon temperature and the total temperature become almost equal, which means the energy distribution is at the end, therefore the saturation in entropy is evident.

Figure 5(a) depicts the relationship between the rate of change of entropy ($\dot{S}$) and time ($t$) with a value of $\tau$ equal to 0.9 and $b$ equal to 0.1. The variation of $\dot{S}$, first exhibits a significantly negative value, then rises and attains a positive value, and eventually reaches zero at different times for different values of $k$. When $k=-1$, there is a prominent positive peak at the beginning of the time period. In all instances, it converges to zero after a certain period of time. The reliability of GSLT was compromised during the first stages of its development. This implies that the larger value of $\tau$ has distinct physical implications for the validity of the GSLT. Figure 5(b) represents the same for $\tau=0.9$ and $b=0.9$. Here, $\dot{S}$ shows similar behavior as of Fig. 5(a) for $k=0$ and $-1$, But for $k=1$, $\dot{S}$ starts from a positive value, then after reaching a negative minima (at $t\approx 10$) it becomes zero. 

The above graph analysis using the GSLT (\ref{49}) exhibited some differences in how the rate of change of entropy ($\dot{S}$) behaved near the beginning of the universe. These differences can be explained below:

The GSLT states that a homogeneous and isotropic system (such as the entire universe) will always grow in entropy. This rule allows for local or temporary subsystem entropy reductions. Entropy increases generally in the cosmos, although particular components may decrease. For an example of a local system, when water freezes to ice, its entropy falls because it has fewer configurations. The emission of heat during this process increases the universe's entropy more than water's. Thus, negative entropy changes in specific regions do not violate the Second Law because other processes compensate for them. We may say that understanding these changes in entropy is crucial in cosmology, since it enables us to accurately simulate the progression of the cosmos, including phase transitions, particle generation, and energy transfer. Also, it may emphasize the complex interaction of thermodynamics, gravitation, and cosmic development. Recall that entropy is a measure of disorder or unpredictability in the cosmos, and its behavior may reflect the intricate dynamics of numerous physical processes. In the case of our study, this phenomenon may arise from the presence of an interaction between normal gravity and a scalar field, resulting in the existence of negative entropy rates. The negative nature of $\dot{S}$ has been achieved in various articles, e.g., in the case of the quintessence field~\cite{Setare}, in $f(T)$ gravity~\cite{Ghosh} and in the scalar field (phantom)~\cite{Chattopadhyay}. In \cite{Nojiri_2004pf} the authors have studied the application of negative entropy (or negative temperature) occurrence in phantom thermodynamics. According to \cite{Nojiri_2004pf, Hawking2,Lineweaver}, we may state that the universe started with a very high or low entropy. If we consider the cosmic microwave background (CMB), which represents a homogeneous, isotropic, isothermal, isobaric, iso-everything universe, the entropy should be maximum. But if so, then how does the universe proceed further? The entropy should then start at a lower value~\cite{Lineweaver}. As time evolves, the availability of phase space increases tremendously due to inflation, and the entropy or disorderliness increase~\cite{Hawking2}. After inflation, the rate of expansion decreases until accelerated expansion comes into play. In our graphs, we also see a sharp variation in the early stages of time, which may be the indication of the rigorous change of entropy as stated in~\cite{Hawking2}. Again, according to~\cite{Lineweaver} if every source of free energy in the universe has already been found, the universe will eventually go to ``heat death'', dissipate all of its free energy, and become less and less complex due to the universe's expansion speeding up and approaching a vacuum state leaving the universe with constant entropy ($\dot{S}=0$). This has also been observed in our graphs of $\dot{S}$.

\subsection{Comparison with other modified theories}
Comprehensive studies of thermodynamics have been done in several modified theories such as $f(R)$ theory \cite{Akbar,Bamba,Wu,Bamba3,Zheng}, Gauss-Bonnet theories \cite{Cai,Akbar2}, scalar-tensor gravity theories \cite{Wu}. The fundamental difference between these theories and our theory is in its geometry. In our model, the K-essence emergent gravity metric ($\bar{G}_{\mu\nu}$) is disformal with the usual gravitational metric ($g_{\mu\nu}$). The new $f(\R,\T)$ gravity model \cite{Panda3} is constructed based on the K-essence theory. Also, in our $f(\R,\T)$ theory, the modified field Eq. (\ref{15}) is different from the usual $f(R,T)$ theory via the modified $\R$ and $\T$. The thermodynamic variables like the surface gravity ($\k_{sg}$), horizon temperature ($T_{h}$), Wald entropy ($S_{h}$), Misner-Sharp energy ($E$) include the $(\K)$ term in their forms. The dynamical behavior of the kinetic component of the scalar field ($\K$) dynamically changes the above-mentioned variables and the first as well as the second laws of thermodynamics. More specifically, this particular field enables us to study the variation of thermodynamical properties with time extensively with the consideration of Eq. (\ref{56}). Let us now compare our results with some of those from standard modified gravity theories in the context of our work.

\subsubsection{Comparison with  $f(R)$ gravity}

Thermodynamics in usual $f(R)$ theory using FLRW metric has been studied in \cite{Akbar,Akbar2}. The expression for the radius of the apparent horizon is given by
\ben
\tilde{r}_{A}=\frac{1}{\sqrt{H^2+\frac{k}{a^2}}}.
\label{59}
\een
Our expression for the radius of the apparent horizon is given in Eq. (\ref{32}). Comparing Eqs. (\ref{32}) and ({\ref{58}}), it can be easily shown that the value of the apparent horizon in our case decreases in the presence of $\K$ as $0<\K<1$. Moreover, the time-dependent terms in Eqs. (\ref{59}) is the Hubble parameter ($H$) and the scale factor ($a$), but in the case of Eq. (\ref{32}) there is an additional time-dependent parameter which is $\K$.\\
The expression of the surface gravity in usual $f(R)$ gravity \cite{Akbar} ($\k_{sg}$) has been given as:
\ben
\k_{sg}=\frac{1}{\tilde{r}_{A}}(1-\frac{\dot{\tilde{r}}_{A}}{2H\tilde{r}_{A}})=-\frac{\tilde{r}_{A}}{2}(2H^2+\dot{H}+\frac{k}{a^2}),
\label{60}
\een
whereas the value of $\k_{sg}$ in our case is given by Eq. (\ref{33}). Though the authors \cite{Akbar}, use ($-,+,+,+$) signature in their article, we change Eq. (\ref{60}), considering the ($+,-,-,-$) signature to compare the results of ours. It is clear that this expression also differs a lot in the presence of $\K$ in our case. The horizon entropy for $f(R)$ gravity has been defined in \cite{Cognola} as:
\ben
S=\frac{Af'(R)}{4G}
\label{61}
\een
which is Eq. (\ref{37}) in our case. The fundamental difference between Eqs. (\ref{37}) and (\ref{61}) lies in the inclusion of the term $f_{\T}$, with $\T$ being formulated based on the K-essence emergent metric (\ref{8}), which incorporates K-essence scalar fields. The modifications in the first law of thermodynamics (\ref{47}) and the generalized second law of thermodynamics (\ref{51}) can also be strongly affected by the existence of K-essence scalar fields.

\subsubsection{Comparison with usual $f(R,T)$ theory}

The differences between the thermodynamics of usual $f(R,T)$ gravity and the same for non-canonical $f(\R,\T)$ are briefly provided below.\\ It is already mentioned in the Introduction that non-canonical $f(\R,\T)$ gravity may be preferable over canonical $f(R,T)$ gravity when it comes to the consideration of the additive form of the function. Moreover, the field equation (\ref{15}), the definition of the energy-momentum tensor (\ref{14}), the expression for $\bar{\Theta}_{\mu\nu}$ (\ref{17}) and the condition of the conservation of energy-momentum tensor (\ref{18}) are different from the usual $f(R,T)$ theory \cite{Harko,Barrientos}. To compare the thermodynamical properties in the usual $f(R,T)$ theory and modified $f(\R,\T)$ theory a graphical analysis in terms of surface gravity has been done in Figs. (\ref{Fig1})--(\ref{Fig3}). We would like to add some more discussions in this subsection.\\
The preceding analysis derived from Eqs. (\ref{59}) and (\ref{60}) is relevant in this scenario as well. The reason lies in Eqs. (\ref{59}) and (\ref{60}), which represent the radius of the apparent horizon and the surface gravity within the conventional FLRW metric. In contrast, Eqs. (\ref{32}) and (\ref{33}) correspond to the same quantities in the K-essence FLRW metric. The equations (\ref{32}), (\ref{33}), as well as (\ref{59}) and (\ref{60}), exhibit independence from the specific formulation of $f(R)$, $f(R,T)$ or $f(\R,\T)$ theories. \\
The first law of thermodynamics in usual $f(R,T)$ gravity has the expression as \cite{Sharif3}
\ben
T_{h}dS_{h}'+T_{h}d_{j}S_{h}'=-dE'+W'dV,
\label{62}
\een
where 
\ben
d_{j}S_{h}'=-\frac{\tilde{r}_{A}}{2GT_{h}}(1+2\pi\tilde{r}_{A}T_{h})d(\frac{f_{R}}{\mathcal{F}}).
\label{63}
\een
The GSLT has been expressed as 
\ben
\dot{S}_{h}'+d_{j}\dot{S}_{h}'+\dot{S}_{tot}'=\frac{24\pi \Xi}{\tilde{r}_{A}bR}\geq 0
\label{64}
\een
With 
\ben
\Xi=(1-b)\dot{\rho}'_{tot}V+(1-\frac{b}{2})(\rho_{tot}'+p_{tot}')\dot{V}
\label{65}
\een
where $R$ is Ricci scalar in usual geometry. The prime ($'$) has been used in all equations from Eq. (\ref{62}) to Eq. (\ref{65}) to signify the parameters distinguishing the original $f(R,T)$ scenario, as described in \cite{Sharif3}.
In our case, the first law of thermodynamics has been achieved in Eq. (\ref{47}) and the expression of $d_{j}S_{h}$ has been shown in Eq. (\ref{48}) whereas the GSLT has been expressed in Eq. (\ref{51}). By comparing Eqs. (\ref{47}) and (\ref{62}), it can be seen that each term has been modified by the presence of $\K$ term. Firstly, there is a factor ($1-\K$) multiplied on the left-hand side of the Eq. (\ref{47}). Secondly, the terms $T_{h}dS_{h}$ (\ref{40}), $d_{j}S_{h}$ (\ref{48}), $dE'$ (\ref{45}), $W'$ (\ref{46}), $\Xi$ (\ref{52}) and $\mathcal{R}$ (\ref{53}) have different structural form due to the presence of the kinetic term ($\K$). The Ricci scalar ($R$) in Eq. (\ref{64}) has been modified into $\mathcal{R}$ in (\ref{51}) which has been expressed in Eq. (\ref{53}). Note that, as $\K\neq 0$ rather than $0<\K<1$, the extra modification terms contribute effectively to the thermodynamical equations. Thirdly, the definition of Ricci scalar ($\R$) and trace of energy-momentum tensor ($\T$) in our modified theory is different from the usual Ricci scalar ($R$) and trace of energy-momentum tensor ($T$) which build the foundation of modified gravity theory.

It should be noted that here we choose to compare our theories in detail on the basis of the usual $f(R)$ and $f(R,T)$ theory. The comparison with other modified theories like scalar-tensor gravity~\cite{Cai,Bamba}, Lovelock gravity~\cite{Cai,Cai3}, Braneworld gravity~\cite{Sheykhi}, nonlinear gravity~\cite{Bamba1} and Gauss-Bonnet gravity \cite{Akbar2} have not been considered in this article as they differ fundamentally in their action and field equation from our theory.

\section{Conclusion}

This article investigates the thermodynamic properties of a homogeneous and isotropic universe using the FLRW model in modified $f(\R,\T)$ gravity within the K-essence geometry. The results differ significantly from those obtained using conventional $f(R,T)$ gravity in the usual FLRW universe framework. 

From the field equations of $f(\R,\T)$ gravity we expressed the semi-conservation equation of this theory in Eq. (\ref{30}) where the total energy exchange term ($\bar{q}_{tot}$) has been changed significantly from the usual $f(R,T)$ case due to the presence of $\K$. Thermodynamic variables such as radius of the apparent horizon ($\r_{A}$) in Eq. (\ref{32}) and surface gravity ($\k_{sg}$) in Eq. (\ref{33}) has also been modified in our case of $f(\R,\T)$ gravity. From the graphs (Figs. \ref{Fig1}-\ref{Fig3}) it is evident that the surface gravity in modified FLRW via $f(\R,\T)$ gravity has distinct features from that of the usual FLRW-based $f(R,T)$ gravity. We get sharp variations of $\k_{sg}$ at the early time ($t\approx 0$) for exponential scale factor (Eq. \ref{34}) over the constant $\k_{sg}$ in the usual case (Figs. 1(a) -- 3(a)). This may be the effect of the scalar field perturbation at the time of inflation \cite{Mukhanov1,Panda5}. For the power law scale factor (\ref{35}) the variation of $\k_{sg}$ in the usual case and modified case shows a similar decreasing nature, but smoothness in the usual case is not present in our case (Figs. 1(b) -- 3(b)). Firstly, at the initial time ($t<1$) we have not achieved any value of $\k_{sg}$ whereas in the usual case, there exists a definite positive value of $\k_{sg}$. Secondly, the value of $\k_{sg}$ has a sharp decrease at the starting point, which may be an indication of the presence of fluctuation at an early time. 

The first law of thermodynamics and the GSLT hold the form of Eqs. (\ref{47}) and (\ref{51}) respectively. The components in expressions such as Wald entropy ($S_{h}$), total energy ($E$), work function ($W'$), extra entropy $(d_{j}S_{h})$, $\Xi$, and $\mathcal{R}$ have different forms from the usual case of $f(R,T)$. For example, to verify the GSLT, considering the Starobinsky type functional form of $f(\R,\T)$ and exponential form of $\K$, we plotted the variation of entropy ($\dot{S}$) with cosmological time ($t$) (Figs. \ref{Fig.4} and \ref{Fig.5}). The plots show the validity or non-validity of GSLT in the context of modified $f(\R,\T)$ gravity. The change in entropy of the early universe may be positive or negative, subject to the value of curvature ($k=-1,0~\text{or}~1$). The negative range of $\dot{S}$ may seem like a violation of the GSLT, but this may happen in local case scenarios. The change in entropy also shows an increasing or decreasing nature as time evolves, depending on the different curvatures of the universe. All of the plots suggest that the entropy of the universe should come to a saturation point at a later time, which leads to the entropy variation decreasing to zero. Also, the greater value of the co-efficient $b$ promotes entropy saturation at an earlier time. Note that $b$ is the coefficient that relates the total temperature $(T_{tot})$ to the horizon temperature ($T_{h}$). 

This investigation presents intriguing features regarding the thermodynamic variables and the validity of the GSLT with regard to the early cosmos. The study of the thermal properties of the cosmos within the context of modified $f(\R,\T)$ gravity, which is an interacting model, may provide a new avenue to understand the present scenario of the universe. The findings suggest that further investigation of scalar field perturbation during the early stages (inflationary scenario) of the cosmos is necessary. However, the study of scalar-field perturbations is beyond the scope of this article. \\

{\bf Acknowledgment:}
A.P. and G.M. acknowledge DSTB, Government of West Bengal, India, for financial support through Grant Nos. 856(Sanc.)/STBT-11012(26)/6/2021-ST SEC dated 3 November 2023. The research by M.K. was carried out in Southern Federal University with financial support of the Ministry of Science and Higher Education of the Russian Federation (State contract GZ0110/23-10-IF). SR sincerely thanks the facilities provided by ICARD, Pune at CCASS, GLA University, Mathura.  \\

{\bf Conflicts of interest:} The authors declare no conflicts of interest.\\

{\bf Data availability:} There is no associated data with this article, and as such, no new data was generated or analyzed in support of this research.\\

{\bf Declaration of competing interest:}
The authors declare that they have no known competing financial interests or personal relationships that could have appeared to influence the work reported in this paper.\\

{\bf Declaration of generative AI in scientific writing:} The authors state that they do not support the use of AI tools to analyze and extract insights from data as part of the study process.\\

\appendix

\section{Representation of $\dot{S}$ in terms of $N$ and $D$}\label{Appendix-A}
\small{
\ben
N&&=3 e^{-\frac{4 t}{\tau }} \pi  \Big(12 e^{t/\tau } k \tau ^2+4 e^{\frac{3 t}{\tau }} k \tau ^2-4 k \tau ^2+e^{\frac{2 t}{\tau }} \Big(\sqrt[3]{-1+e^{t/\tau }}-12 k \tau ^2\Big)\Big) \Big(-b \Big(1-e^{-\frac{t}{\tau }}\Big) \Big(\Big(-2+7 e^{-\frac{t}{\tau }}\Big) \alpha\nonumber\\
&&-3 e^{-\frac{3 t}{\tau }} \Big(-1+e^{t/\tau }\Big)^3 \tau ^2\Big) \Big(-72 e^{t/\tau } k \tau ^2 \Big(-1+e^{t/\tau }\Big)^{2/3}+36 e^{\frac{2 t}{\tau }} k \tau ^2 \Big(-1+e^{t/\tau }\Big)^{2/3}\nonumber\\
&&+36 k \tau ^2 \Big(-1+e^{t/\tau }\Big)^{2/3}+e^{\frac{3 t}{\tau }}\Big) \Big(-2 e^{-\frac{t}{\tau }} \Big(-3+e^{t/\tau }\Big) \nonumber\\
&&\sqrt{\frac{\Big(-1+e^{t/\tau }\Big)^3 \tau ^2}{-72 e^{t/\tau } k \tau ^2 \Big(-1+e^{t/\tau }\Big)^{2/3}+36 e^{\frac{2 t}{\tau }} k \tau ^2 \Big(-1+e^{t/\tau }\Big)^{2/3}+36 k \tau ^2 \Big(-1+e^{t/\tau }\Big)^{2/3}+e^{\frac{3 t}{\tau }}}} \sqrt[3]{-1+e^{t/\tau }}\nonumber\\
&&+3 e^{-\frac{2 t}{\tau }} \tau  \sqrt{\frac{e^{\frac{3 t}{\tau }} \Big(-1+e^{t/\tau }\Big)^{2/3}}{108 e^{t/\tau } k \tau ^2-108 e^{\frac{2 t}{\tau }} k \tau ^2-36 k \tau ^2+e^{\frac{3 t}{\tau }} \Big(\sqrt[3]{-1+e^{t/\tau }}+36 k \tau ^2\Big)}} \Big(-1+e^{t/\tau }\Big)\nonumber\\
&&+36 \Big(1-e^{-\frac{t}{\tau }}\Big)^3 k \tau ^2 \sqrt{\frac{\Big(-1+e^{t/\tau }\Big)^3 \tau ^2}{-72 e^{t/\tau } k \tau ^2 \Big(-1+e^{t/\tau }\Big)^{2/3}+36 e^{\frac{2 t}{\tau }} k \tau ^2 \Big(-1+e^{t/\tau }\Big)^{2/3}+36 k \tau ^2 \Big(-1+e^{t/\tau }\Big)^{2/3}+e^{\frac{3 t}{\tau }}}}\Big)\nonumber\\
&& \Big(-1+e^{t/\tau }\Big)^{2/3}+16 (b-1) e^{-\frac{7 t}{\tau }} \pi  \tau  \Big(-1944 k \tau ^5+54 e^{t/\tau } k (252 \tau +1) \tau ^4-18 e^{\frac{2 t}{\tau }} k (252 \alpha +\tau  (2268 \tau +17)) \tau ^3\nonumber\\
&&+6 e^{\frac{8 t}{\tau }} k \Big(3 \tau ^2+2 \alpha \Big) \tau ^2+18 e^{\frac{3 t}{\tau }} \Big(3 \tau  \sqrt[3]{-1+e^{t/\tau }}+k \alpha  (1080 \tau -77)+3 k \tau ^2 (1260 \tau +13)\Big) \tau ^2\nonumber\\
&&-3 e^{\frac{4 t}{\tau }} \Big((72 \tau -1) \sqrt[3]{-1+e^{t/\tau }}+270 k \tau ^2 (84 \tau +1)+8 k \alpha  (1350 \tau -169)\Big) \tau ^2+3 e^{\frac{5 t}{\tau }} \Big(42 \alpha  \sqrt[3]{-1+e^{t/\tau }}\nonumber\\
&&+8 k \alpha  \tau  (1080 \tau -161)+3 \tau  \Big(36 \tau  \sqrt[3]{-1+e^{t/\tau }}-\sqrt[3]{-1+e^{t/\tau }}+4536 k \tau ^3+50 k \tau ^2\Big)\Big) \tau +e^{\frac{7 t}{\tau }} \Big(3 \Big(18 \tau  \sqrt[3]{-1+e^{t/\tau }}\nonumber\\
&&-\sqrt[3]{-1+e^{t/\tau }}+648 k \tau ^3-18 k \tau ^2\Big) \tau ^2+\alpha  \Big(36 \tau  \sqrt[3]{-1+e^{t/\tau }}+\sqrt[3]{-1+e^{t/\tau }}+1296 k \tau ^3+66 k \tau ^2\Big)\Big)-\nonumber\\
&&e^{\frac{6 t}{\tau }} \Big(9 \Big(24 \tau  \sqrt[3]{-1+e^{t/\tau }}-\sqrt[3]{-1+e^{t/\tau }}+1512 k \tau ^3+6 k \tau ^2\Big) \tau ^2+\alpha  \Big(162 \tau  \sqrt[3]{-1+e^{t/\tau }}-49 \sqrt[3]{-1+e^{t/\tau }}\nonumber\\
&&+9720 k \tau ^3-1116 k \tau ^2\Big)\Big)\Big) \Big(-1+e^{t/\tau }\Big)^{11/3}+48 (b-1) e^{-\frac{7 t}{\tau }} \pi  \tau ^2 \Big(9 e^{t/\tau } \tau ^2-3 \tau ^2+e^{\frac{3 t}{\tau }} \Big(3 \tau ^2+2 \alpha \Big)\nonumber\\
&&-e^{\frac{2 t}{\tau }} \Big(9 \tau ^2+7 \alpha \Big)\Big) \Big(204 e^{t/\tau } k \tau ^2-180 e^{\frac{2 t}{\tau }} k \tau ^2-72 k \tau ^2+e^{\frac{4 t}{\tau }} \Big(\sqrt[3]{-1+e^{t/\tau }}+12 k \tau ^2\Big)-4 e^{\frac{3 t}{\tau }} \Big(\sqrt[3]{-1+e^{t/\tau }}-9 k \tau ^2\Big)\Big)\nonumber\\
&& \Big(-1+e^{t/\tau }\Big)^{14/3}-48 e^{-\frac{7 t}{\tau }} \Big(b \Big(-2+e^{\frac{2 t}{\tau }}\Big)+e^{t/\tau }-e^{\frac{2 t}{\tau }}\Big) \pi  \tau ^2 \Big(-72 e^{t/\tau } k \tau ^2 \Big(-1+e^{t/\tau }\Big)^{2/3}\nonumber\\
&&+36 e^{\frac{2 t}{\tau }} k \tau ^2 \Big(-1+e^{t/\tau }\Big)^{2/3}+36 k \tau ^2 \Big(-1+e^{t/\tau }\Big)^{2/3}+e^{\frac{3 t}{\tau }}\Big) \Big(9 e^{t/\tau } \tau ^2-3 \tau ^2+e^{\frac{3 t}{\tau }} \Big(3 \tau ^2+2 \alpha \Big)-\nonumber\\
&&e^{\frac{2 t}{\tau }} \Big(9 \tau ^2+7 \alpha \Big)\Big) \Big(-1+e^{t/\tau }\Big)^4\Big)
\label{A1}
\een}

\vspace{1.75in}

{\small
\ben
D&&=b \Big(1-e^{-\frac{t}{\tau }}\Big)^{17/2} \Big(-1+e^{t/\tau }\Big)^{13/3} (\lambda +1) \tau ^9 \sqrt{\frac{\Big(-1+e^{t/\tau }\Big)^{10/3} \tau ^2}{108 e^{t/\tau } k \tau ^2-108 e^{\frac{2 t}{\tau }} k \tau ^2-36 k \tau ^2+e^{\frac{3 t}{\tau }} \Big(\sqrt[3]{-1+e^{t/\tau }}+36 k \tau ^2\Big)}}\nonumber\\
&& \Big(\frac{e^{-\frac{t}{\tau }} \Big(108 e^{t/\tau } k \tau ^2-108 e^{\frac{2 t}{\tau }} k \tau ^2-36 k \tau ^2+e^{\frac{3 t}{\tau }} \Big(\sqrt[3]{-1+e^{t/\tau }}+36 k \tau ^2\Big)\Big)}{\Big(-1+e^{t/\tau }\Big)^{7/3} \tau ^2}\Big)^{5/2} \Big(\frac{e^{\frac{2 t}{\tau }} \Big(-7+2 e^{t/\tau }\Big)}{6 \Big(-1+e^{t/\tau }\Big)^3 \tau ^2}+\frac{12 k}{\sqrt[3]{-1+e^{t/\tau }}}\nonumber\\
&&+\frac{e^{t/\tau } \sqrt{\frac{e^{\frac{3 t}{\tau }} \Big(-1+e^{t/\tau }\Big)^{2/3}}{108 e^{t/\tau } k \tau ^2-108 e^{\frac{2 t}{\tau }} k \tau ^2-36 k \tau ^2+e^{\frac{3 t}{\tau }} \Big(\sqrt[3]{-1+e^{t/\tau }}+36 k \tau ^2\Big)}}}{2 \Big(-1+e^{t/\tau }\Big)^{7/3} \tau  \sqrt{\frac{\Big(-1+e^{t/\tau }\Big)^3 \tau ^2}{-72 e^{t/\tau } k \tau ^2 \Big(-1+e^{t/\tau }\Big)^{2/3}+36 e^{\frac{2 t}{\tau }} k \tau ^2 \Big(-1+e^{t/\tau }\Big)^{2/3}+36 k \tau ^2 \Big(-1+e^{t/\tau }\Big)^{2/3}+e^{\frac{3 t}{\tau }}}}}-\nonumber\\
&&\frac{e^{\frac{2 t}{\tau }}}{2 \Big(-1+e^{t/\tau }\Big)^3 \tau ^2}\Big)
\label{A2}
\een}


\end{document}